\documentclass{article}
\usepackage{cite,amsmath,amssymb,color,graphicx,subfigure,fullpage,upgreek, epstopdf}
\usepackage[hidelinks,pdfstartview=FitH,bookmarksopen=true,bookmarksopenlevel=4,pdfpagemode=UseOutlines]{hyperref}
\usepackage{authblk}

\begin{document}

\title{A Technical Primer on the Physical Modeling of Diffusion-Encoded Magnetic Resonance Experiments: A Random Process Perspective}

\author{Justin P. Haldar}
\affil{Signal and Image Processing Institute,\\ Ming Hsieh Department of Electrical and Computer Engineering, \\ University of Southern California}

\maketitle

\begin{abstract}
 Diffusion-encoded magnetic resonance (MR) experiments can provide important insights into the microstructural characteristics of a variety of biological tissues and other fluid- or gas-filled porous media.  The physics of diffusion encoding has been studied extensively over the span of many decades, and many excellent descriptions can be found in the literature -- see, e.g., Refs.~\cite{callaghan1991, lebihan1995,jones2010, mori2013, johansen-berg2014}.  However, many of these descriptions spend relatively little time focusing on random process descriptions of the diffusion process, instead relying on different abstractions.   In this primer, we describe diffusion-encoded MR experiments from a random process perspective. While the results we derive from this perspective are quite standard (and match the results obtained with other arguments), we expect that the alternative derivations may be insightful for some readers.  This primer is intended for technical readers who have a graduate-level understanding of random processes.  Readers are also expected to already have good familiarity with the basics of MR, and we anticipate that a signal processing perspective on MR \cite{liang2000} will be especially complementary to the random process perspectives presented herein.
\end{abstract}

\section{Introduction}
\begin{figure}[btp]
\centering
\includegraphics[width=0.24\linewidth]{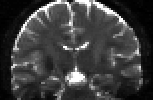}
\includegraphics[width=0.24\linewidth]{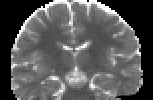}
\includegraphics[width=0.24\linewidth]{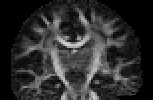}
\includegraphics[width=0.24\linewidth]{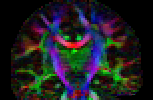}\\[0.25em]
\subfigure[Anatomical Image]{\includegraphics[width=0.24\linewidth]{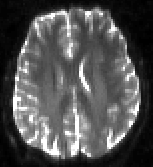}}
\subfigure[Mean Diffusivity]{\includegraphics[width=0.24\linewidth]{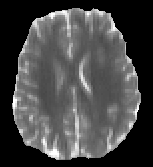}}
\subfigure[Fractional Anisotropy]{\includegraphics[width=0.24\linewidth]{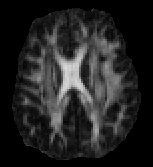}}
\subfigure[Orientation]{\includegraphics[width=0.24\linewidth]{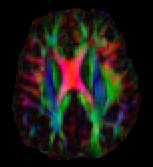}}
\subfigure[Orientation Distributions]{\includegraphics[height=3in]{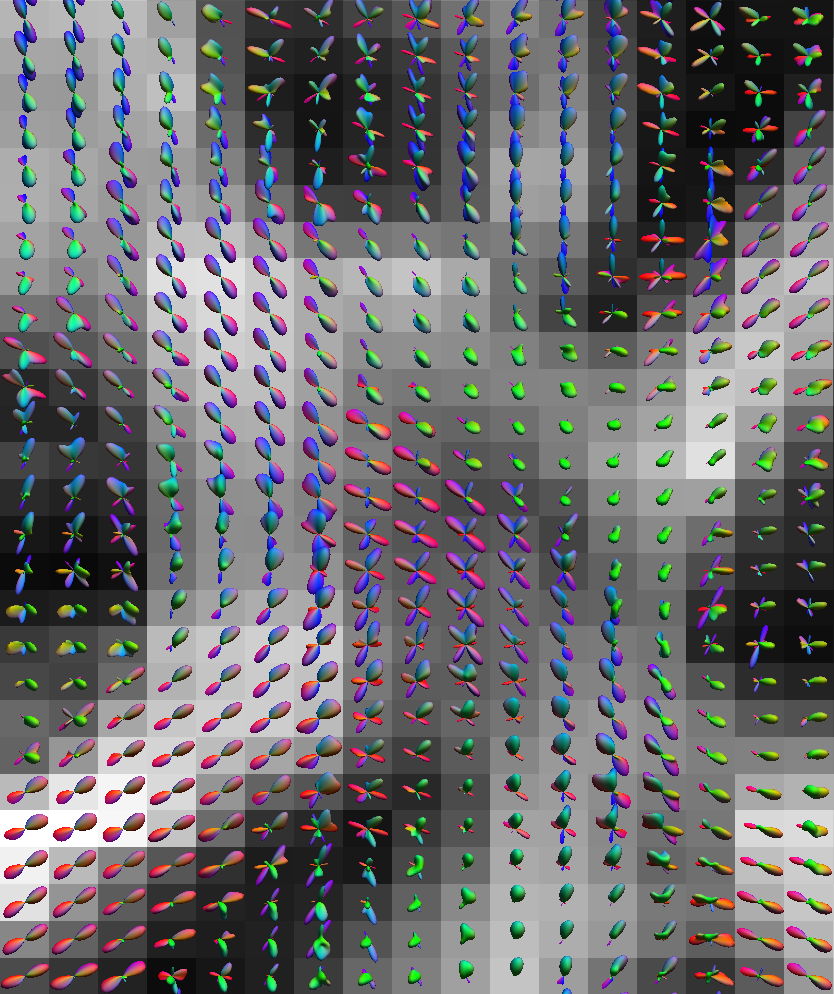}}
\subfigure[Tractography]{\includegraphics[height=3in]{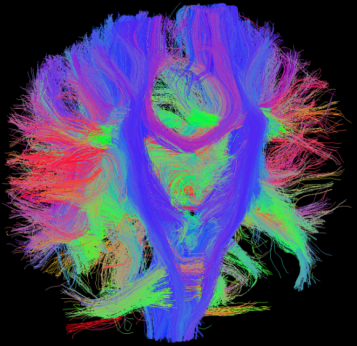}}
 \caption{While (a) conventional anatomical images ($T_2$-weighted in this case) can show very useful information about the macroscopic (millimeter-scale) features of brain tissue, they provide limited information about features that are smaller than the spatial resolution of the image.  In contrast, various  (b-d)  quantitative parameter maps extracted from diffusion MR measurements provide information about the microscopic (micron-scale) features of the tissue by identifying the  characteristics of the thermally-driven (diffusive) random displacement of water molecules within the tissue over short (microsecond) timescales. The insights provided by quantitative parameter maps like (b) mean diffusivity and (c) fractional anisotropy are sensitive to the structure, organization, and integrity of various microscopic tissue compartments, and can provide valuable biomarkers for subtle tissue changes due to pathology, learning/plasticity, and development/aging. Diffusion MR also provides valuable information about (d,e) tissue orientation, which can be used to produce (f) tractography results that provide detailed insights into the pathways of white-matter fiber bundles within the brain.  The results in this figure were obtained using methods described in \cite{haldar2012,varadarajan2015,haldar2013,bhushan2015} and the BrainSuite Diffusion Pipeline (\url{http://brainsuite.org/processing/diffusion/}).  }
 \label{fig:dmri}
\end{figure}

Diffusion magnetic resonance (MR) is a powerful modality that is sensitive to the microstructural characteristics of biological tissues and other fluid- or gas-filled porous media.  Diffusion MR is particularly utilized in studies of the central nervous system, in which the tissue microstructure is quite intricate and the diffusion signal is sensitive to minute biophysical features that are largely invisible to conventional anatomical MR methods.  Examples are shown in Fig.~\ref{fig:dmri}.

\begin{figure}
\centering
 \includegraphics[width=6in]{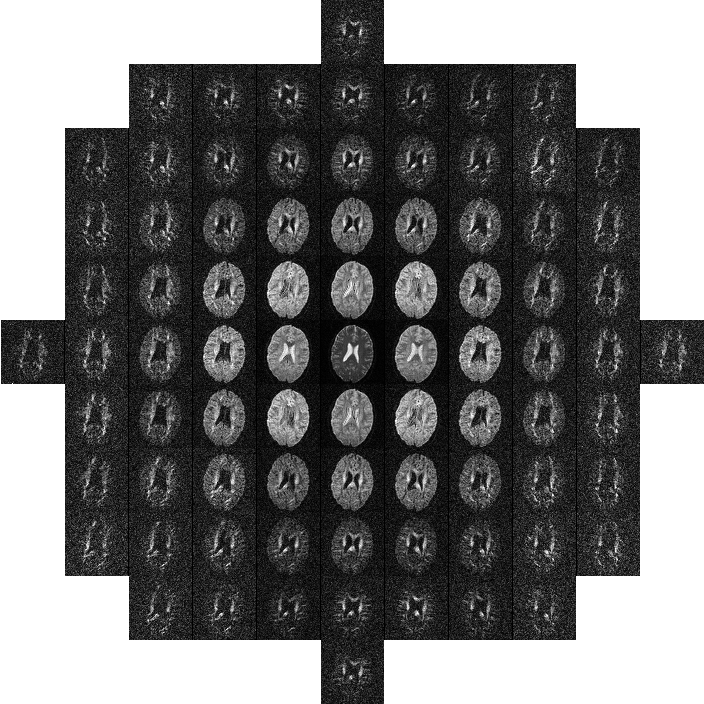} 
 \caption{Illustration of diffusion-weighted images.  The image at the center shows a brain slice with negligible diffusion weighting.  The remaining images show the exact same slice, but with increasing amounts of diffusion weighting as we move further from the center along different orientations.  These images have been normalized so that the features are easier to visualize, but the MR signal generally decays with increasing amounts of diffusion weighting.  In addition, diffusion weighting is often an oriented (directional) quantity. For the data shown in this figure, the horizontal axis corresponds to different amounts of diffusion weighting along the left-right orientation, while the vertical axis corresponds to different amounts of diffusion weighting along the anterior-posterior orientation.  If diffusion is anisotropic (e.g., if diffusing particles tend to move farther along one direction than they do along others), then the amount of signal decay will vary as a function of the diffusion encoding  orientation.  In tissues like brain white matter, the diffusion process is highly oriented, which leads to visually-obvious orientation-dependent effects that can be seen in the figure.   }\label{fig:data} 
\end{figure}

Diffusion MR is possible because there are ways of manipulating the parameters of an MR experiment so that the measured signal becomes sensitive to diffusion characteristics.  Specifically, by applying different degrees of ``diffusion weighting,'' we can obtain the type of diffusion-weighted images shown in Fig.~\ref{fig:data}.  The extraction of microstructural information (of the type shown in Fig.~\ref{fig:dmri}) is generally achieved by fitting parametric models of the signal to this kind of diffusion-weighted data.  But in order to develop appropriate signal models, we need to have an understanding of the physics of diffusion encoding -- the main topic of this primer!

\section{Random Process Description of Diffusion Encoding}
The signal measured in an MR experiment originates from the microscopic magnetic moments of an ensemble of atomic nuclei within an excited spatial region of interest. Following common MR terminology \cite{liang2000}, we will  refer to these nuclei as \emph{spins}, in reference to the important quantum mechanical property that gives rise to the MR phenomenon.   While there are many MR-active nuclei that can be measured, the $^1$H nucleus is by far the most common and widely used, and our description will focus on the $^1$H case (also called \emph{proton} MR because the $^1$H nucleus is simply a single proton) and the diffusion of water molecules (i.e., H$_2$O molecules, which each contain two hydrogen nuclei).  

In what follows, we will describe diffusion MR from the perspective of classical physics, ignoring the fact that the MR phenomenon is inherently quantum mechanical in nature.  There are good reasons for invoking a classical description \cite{hanson2008}, e.g., the classical description is much simpler and more intuitive to understand, and the results obtained with a simplified classical description yield essentially the same results as a quantum mechanical description when averaged over a large ensemble of spins.  However, it should be noted that there are major differences in the behavior of individual spins and large ensembles of spins that we are neglecting (e.g., measurement of an individual spin will cause collapse into an eigenstate) -- although our description of diffusion MR will be based on the classical behavior of individual particles (ignoring concepts like the quantum state of the particle and implicitly assuming perfect nuclear polarization), readers should be cautioned that this is a gross simplification, yet still useful as the overall signal characteristics are not changed.

In our description, we will also ignore the details of spatial localization, and simply assume that the signal we measure has been appropriately localized (e.g., using an MR imaging experiment, through spatially-selective excitation, through the use of receiver coils with spatially-localized sensitivity profiles, etc.) -- see, e.g., Ref.~\cite{liang2000} for a detailed explanation of spatial localization in MR experiments.  However, it should be noted that imaging gradients aren't any different from diffusion-weighting gradients, and also contribute diffusion weighting that should not be ignored.

Consider a voxel containing an ensemble of $N_s$ excited spins.  At some initial time $t=0$, the $i$th spin from this ensemble is assumed to be located at a position $\mathbf{x}_i(0) = [x_i(0), y_i(0), z_i(0)]^T$ in three-dimensional space.  We assume that these initial spin locations are drawn i.i.d. from some initial probability distribution $p_{\mathrm{init}}(\mathbf{x})$.  As time increases, the spins experience molecular diffusion (i.e., the position of each spin is expected to evolve randomly  due to thermal agitation), leading to trajectories $\mathbf{x}_i(t)$, $i=1,\ldots,N_s$, that are random processes.  These trajectories are constrained by the local microstructure of the porous media  that the spins are diffusing within (e.g., biological tissue), meaning that the ability to identify the characteristics of these trajectories can enable  unique insights into the microstructural properties of the porous media.  For example, the diffusion signal can provide insight into the orientation of fibrous tissues (like white matter fiber bundles in the brain or muscle fibers in the body), the sizes of cells and/or pores within biological tissues, etc.  In addition, the diffusion signal can be a sensitive marker for changes in the microstructural configuration, and therefore has many potential uses in various application domains where such changes may be important to identify.

\begin{figure}[p] 
\centering 
\subfigure[Static Spins]{\includegraphics[width=0.48\linewidth]{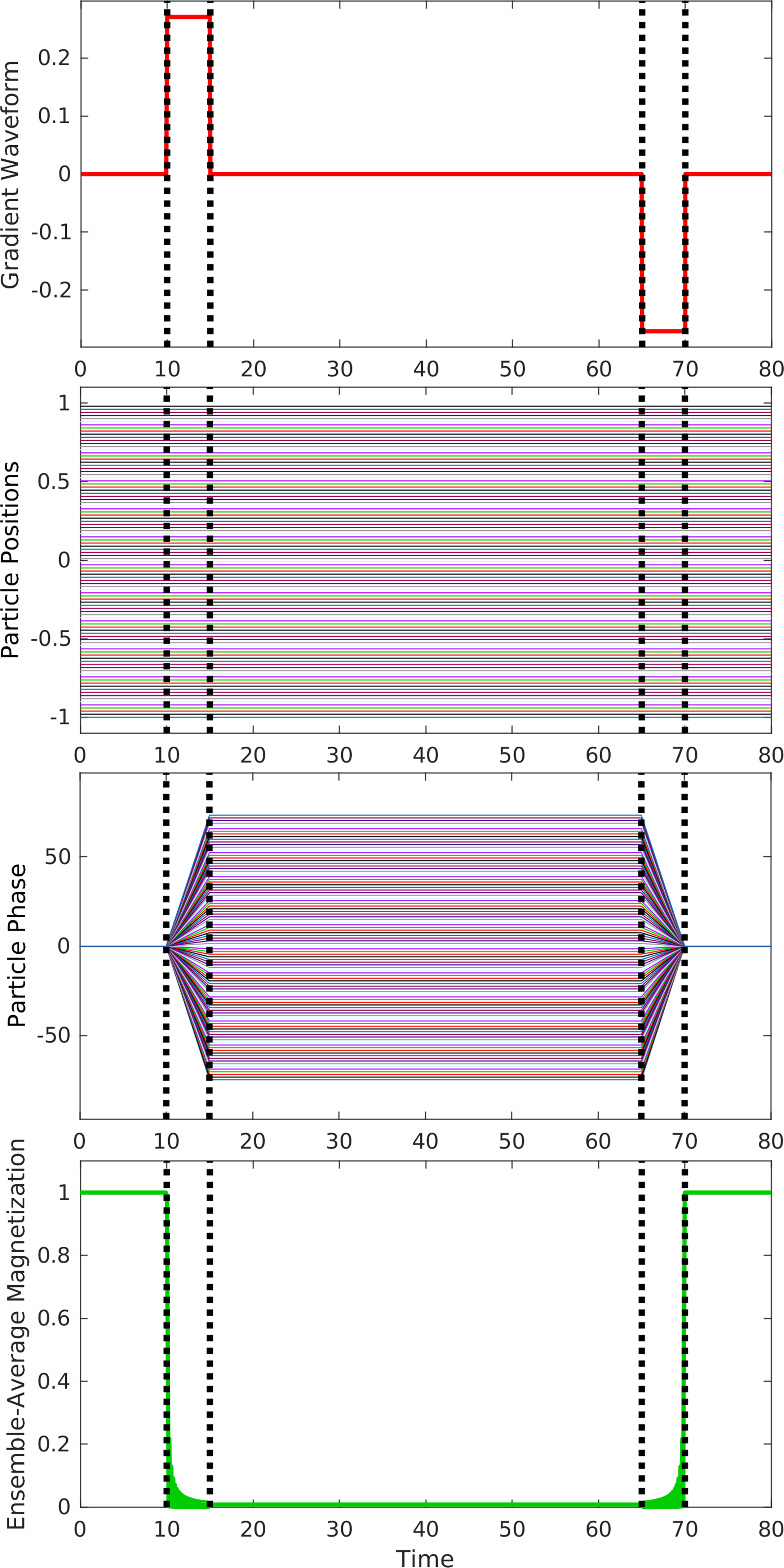}}
\subfigure[Moving Spins]{\includegraphics[width=0.48\linewidth]{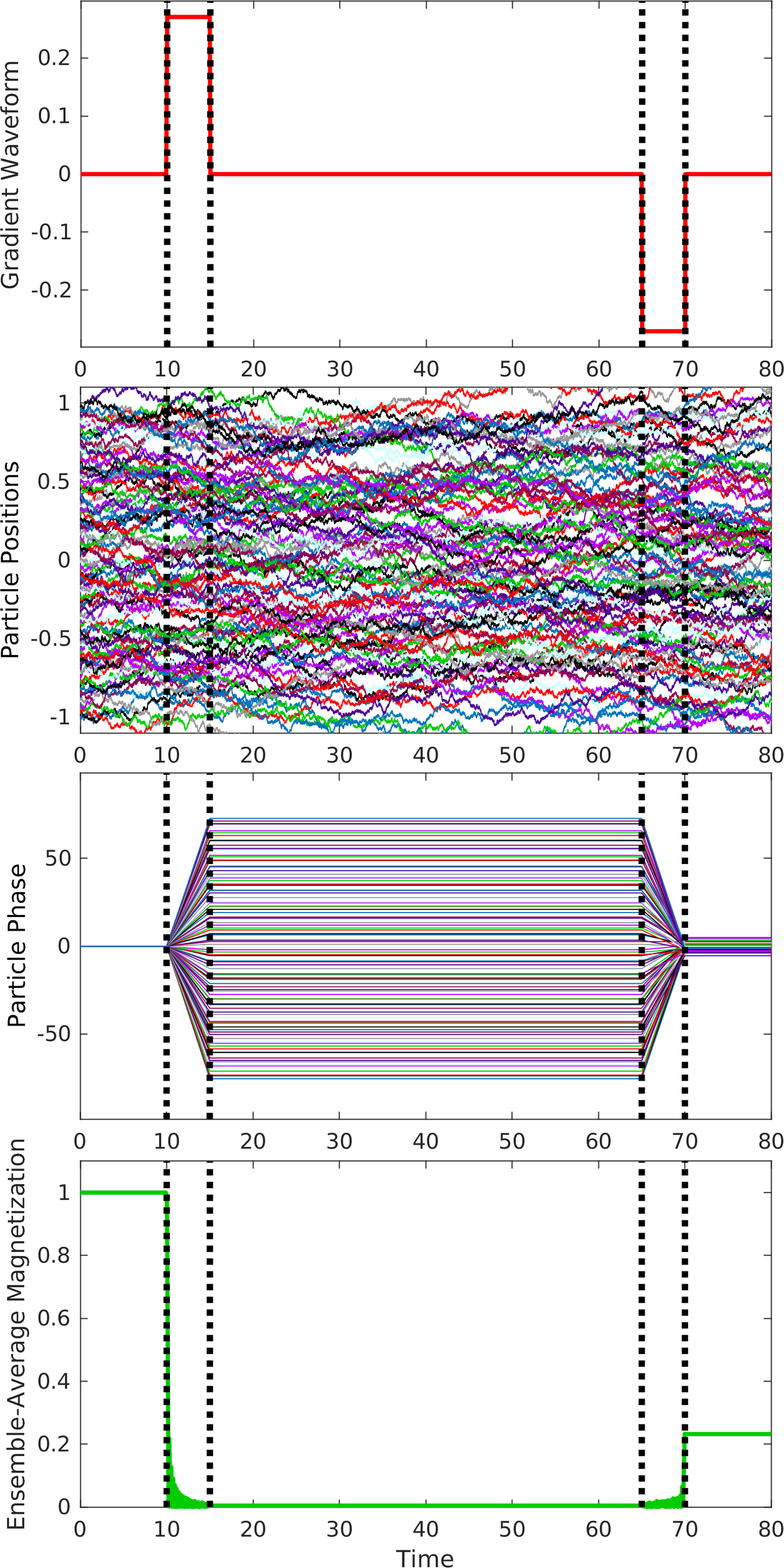}} 
\caption{Illustration of how random thermal motion in the presence of magnetic field gradients can lead to a loss of phase coherence, which ultimately results in attenuation of the measured signal.  (a) The plots on the left correspond to a simulation with static (non-moving) spins in the absence of relaxation.  As can be seen, the spins originally have coherent phase, but the application of the positive gradient pulse causes each spin to accumulate phase based on its position.  Because the spins are each at different locations, this results in phase dispersion and fast decay of the ensemble-average magnetization.  However, this process is completely reversible because the spins do not change position, and the application of a negative gradient pulse allows the accumulated phase to be perfectly rewound.  At the end of this process, the spins regain perfect phase coherence and the ensemble-average magnetization returns to its initial value.  (b) The plots on the right correspond to a simulation with moving spins.  In this case, due to the spin motion, the spins are not at the same locations during the negative gradient pulse as they were during the positive gradient pulse, such that the negative gradient pulse does not perfectly rewind the accumulated phase.  In the end, this leads to residual phase dispersion and a net reduction in the ensemble-average magnetization.}\label{fig:illust}  
\end{figure} 

In a diffusion-encoded MR experiment, a linear gradient field $\mathbf{g}(t) = [g_x(t), g_y(t), g_z(t)]^T$ is turned on while the spins are moving.  Based on the physical principles of MR \cite{liang2000}, this linear gradient field will cause an individual excited  spin that moves according to spatial trajectory $\mathbf{x}(t)$ to accumulate phase according to 
\begin{equation}
 \phi(t) = -\gamma \int_0^t \mathbf{g}(\tau) \cdot \mathbf{x}(\tau) d\tau,\label{eq:phi}
\end{equation}
where $\gamma$ is the gyromagnetic ratio (and equals $2\pi \cdot 42.58$ MHz/T for $^1$H), and we have ignored various real-world effects like relaxation, magnetic field inhomogeneity, gradient nonlinearity, concomitant fields, etc., that can also contribute to phase accrual.     Since each spin  contributes equally to the bulk signal, the signal measured at time $t$ from an ensemble of $N_s$ spins will be proportional to
\begin{equation}
  \sum_{i=1}^{N_s} e^{-i \gamma \int_{0}^t \mathbf{g}(\tau) \cdot \mathbf{x}_i(\tau) d\tau}.\label{eq:sum}
\end{equation}
The fact that different spins accumulate different amounts of phase  as a result of taking different random walks through the environment leads to a loss of phase coherence as illustrated in Fig.~\ref{fig:illust}, ultimately resulting in attenuation of the measured signal as spins with different phase interfere destructively.   If we assume that $N_s$ is large and invoke the law of large numbers, then Eq.~\eqref{eq:sum} will be well-approximated by
\begin{equation}
 N_s \cdot E\left[ e^{-i \gamma \int_{0}^t \mathbf{g}(\tau) \cdot \mathbf{x}(\tau) d\tau} \right],\label{eq:noecho}
\end{equation}
where $E[\cdot]$ denotes statistical expectation.  

The expression above is based on diffusion encoding in the absence of a spin-echo (180$^\circ$) pulse.  However, most diffusion-encoding MR sequences involve a spin-echo pulse, which has the effect of inverting the accumulated phase.  If we assume that an ideal instantaneous (i.e., occupying no time, such that the location of the spins is assumed frozen during the pulse) spin-echo pulse is applied at time $t_s$ (with $0<t_s<t$), then the modified signal model will be proportional to
\begin{equation}
 N_s \cdot E\left[e^{+i \gamma \int_{0}^{t_s} \mathbf{g}(\tau) \cdot \mathbf{x}(\tau) d\tau} e^{-i \gamma \int_{t_s}^t \mathbf{g}(\tau) \cdot \mathbf{x}(\tau) d\tau} \right].\label{eq:withspin}
\end{equation}
This expression is easily generalized to acquisitions involving more than one spin-echo pulse.  We can simplify Eq.~\eqref{eq:withspin} to an expression with a similar form to Eq.~\eqref{eq:noecho}, i.e.,
\begin{equation}
 N_s \cdot E\left[ e^{-i \gamma \int_{0}^t \tilde{\mathbf{g}}(\tau) \cdot \mathbf{x}(\tau) d\tau} \right],\label{eq:noyesecho}
\end{equation}
by defining an ``effective gradient'' $\tilde{\mathbf{g}}(t)$ as 
\begin{equation}
 \tilde{\mathbf{g}}(t) = \left\{ \begin{array}{rl} -\mathbf{g}(t), & 0 \leq t < t_s \\ \mathbf{g}(t), &  t_s < t. \end{array} \right.
\end{equation}
As such, the remainder of the paper will focus on the simplified expression from Eq.~\eqref{eq:noyesecho} with effective gradients, with no need for more complicated notation to accommodate  spin-echoes.  

From a probability perspective, Eq.~\eqref{eq:noyesecho} can be interpreted in terms of the characteristic function of the phase accumulation random variable arising from the interaction between the random trajectory $\mathbf{x}(t)$ and the deterministic effective gradient waveform $\tilde{\mathbf{g}}(t)$.  Calculating the expectation in Eq.~\eqref{eq:noyesecho} can depend on what assumptions we make about the characteristics of the random process $\mathbf{x}(t)$.

\begin{figure}[p]
\centering 
\subfigure[Isotropic Brownian Motion]{\includegraphics[width=0.48\linewidth,clip=]{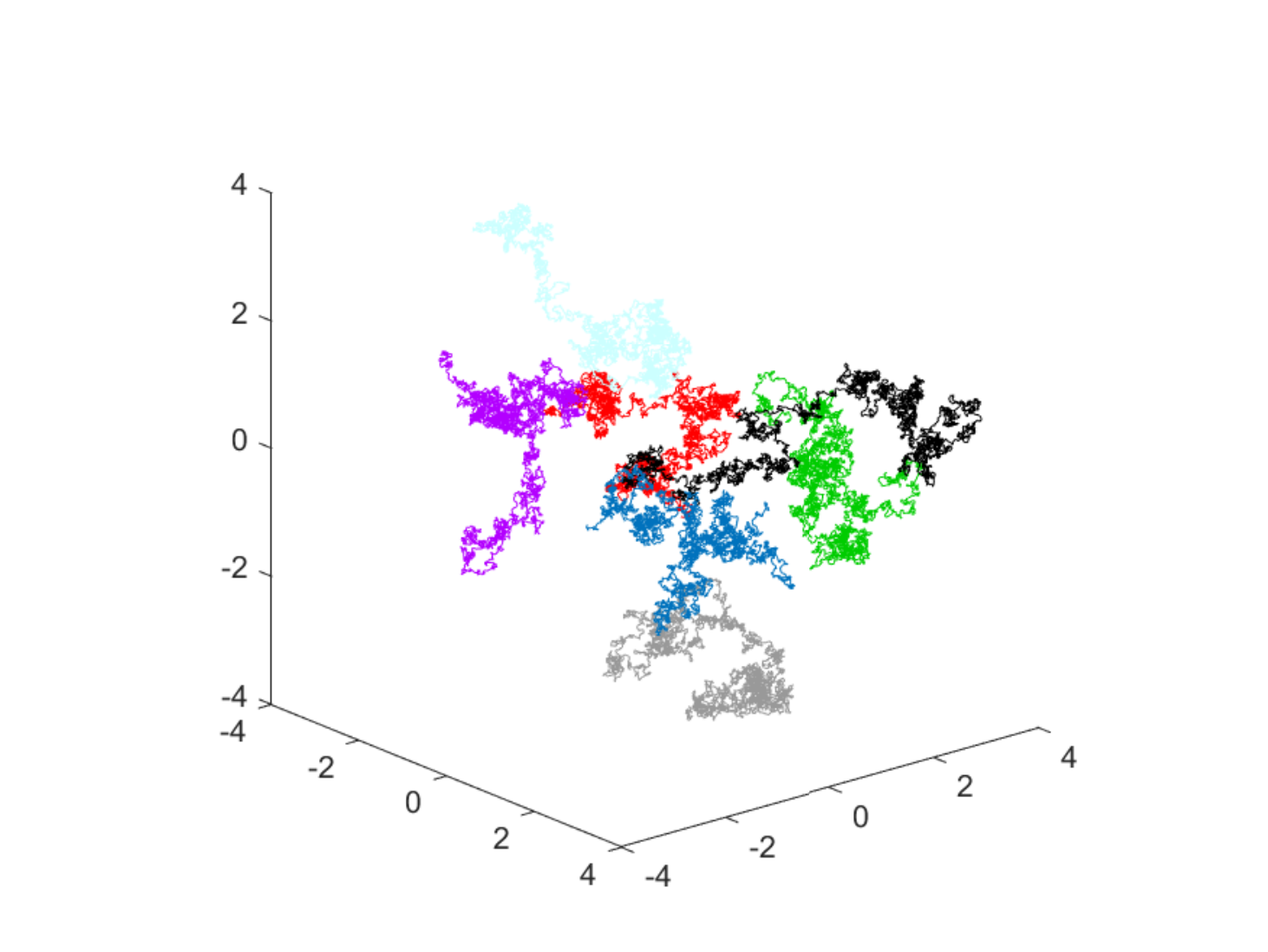}}
\subfigure[Anisotropic Brownian Motion]{\includegraphics[width=0.48\linewidth,clip=]{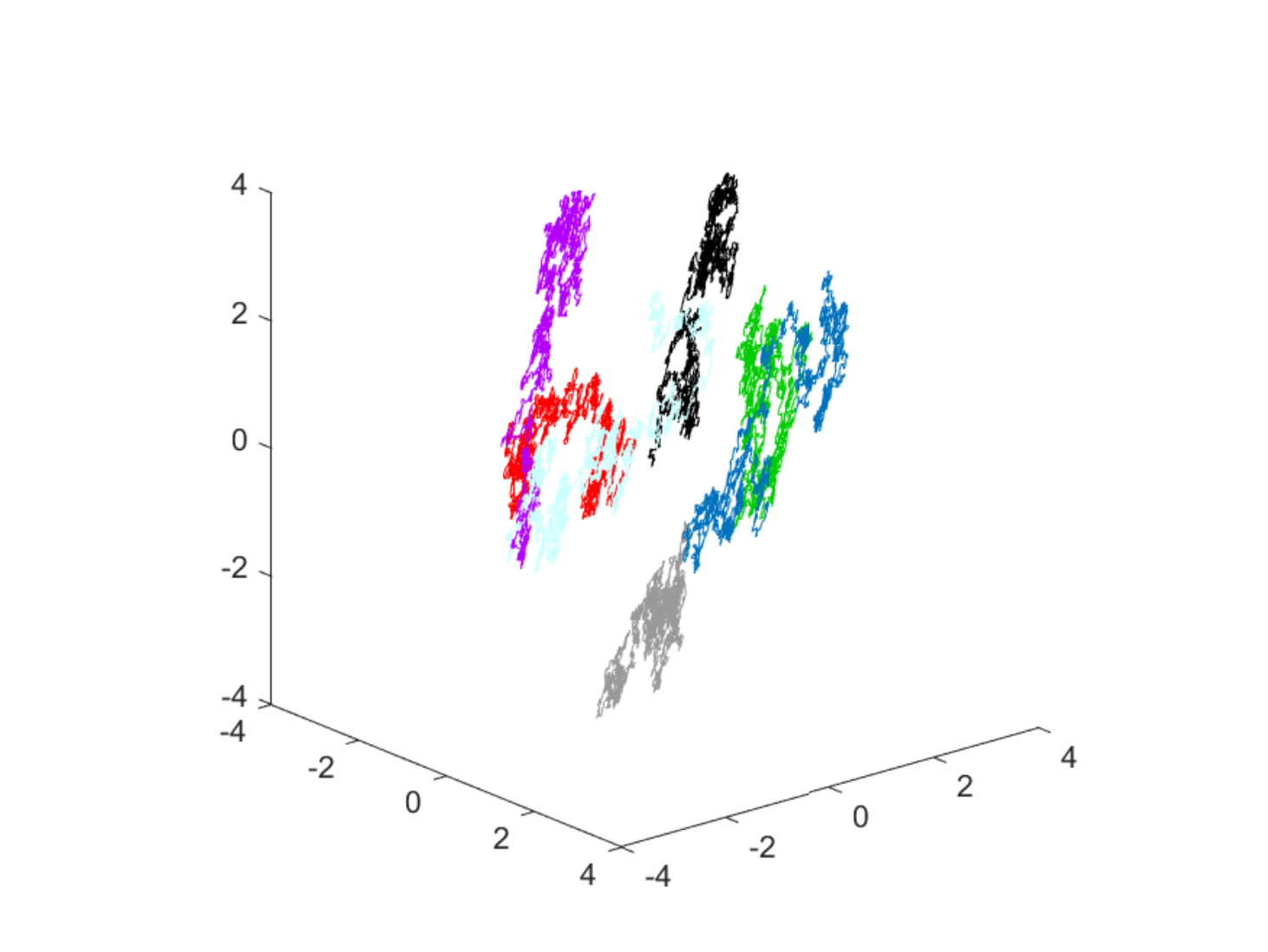}}
\subfigure[Diffusion Constrained by Impermeable Boundaries]{\includegraphics[width=0.48\linewidth,clip=]{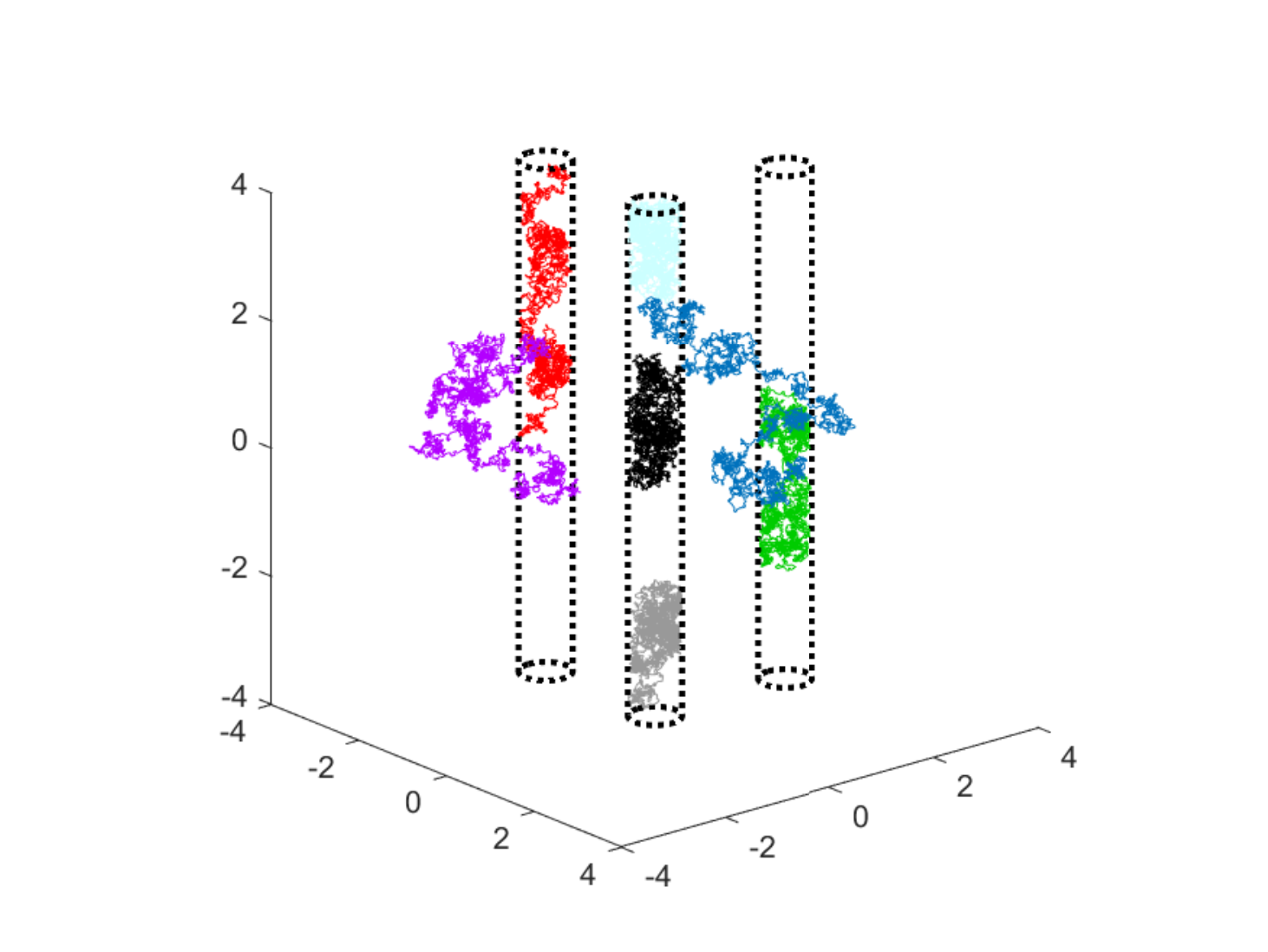}}
\caption{Illustration of random trajectories obtained under different modeling assumptions about the diffusion process.  Each plot illustrates the trajectories of seven particles, where the trajectories of different particles are differentiated using color.  (a) Isotropic Brownian motion, in which particles are able to diffuse freely in a homogeneous environment (without barriers or other restrictions) and there is no preferential displacement direction.  (b) Anisotropic Brownian motion, in which particles are able to diffuse freely in a homogeneous environment (without barriers or other restrictions), but there is still a preferential displacement direction. (c) Diffusion in the presence of impermeable boundaries.  In this case, there are impermeable cylinders that influence different particles in different ways.  The particles inside the cylinders are highly constrained by the cylinder, and as a result, end up having most of their displacement along the z-axis.  The particles outside the cylinder have relatively limited interactions with the cylindrical boundary and follow trajectories that are fairly close to Brownian motion.  This type of model can, e.g., be used to represent tissues like white matter, in which some water molecules are highly constrained by the white matter fiber geometry while other water molecules are somewhat less constrained.  }\label{fig:brown}
\end{figure}

\section{Random Process Models}
\subsection{Isotropic Brownian Motion}
In a very simple system (e.g., an infinitely-large glass of water with no boundaries), one method for modeling molecular diffusion is to assume that it is a 3D isotropic Brownian motion (Wiener process).  An illustration of this case is shown in Fig.~\ref{fig:brown}(a).  A 1D Brownian motion $w(t)$ (Wiener process) is characterized by four properties \cite{hajek2015}:
\begin{enumerate}
 \item $w(0)=0$ with probability 1.
 \item $w(t)$ has independent increments.
 \item $w(t) - w(s)$ is a zero-mean  normal (Gaussian) random variable with variance $\sigma^2 |t-s|$ for any $t \geq 0$ and $s \geq 0$, where $\sigma$ is a parameter of the distribution.
 \item $w(t)$ is a continuous function of $t$ with probability 1.
\end{enumerate}
Readers should be cautioned that the term ``Brownian motion'' is defined differently in random processes than it is in the physical sciences.  In the physical sciences, Brownian motion simply describes the random fluctuations of a particle's position due to thermal agitation (i.e., the phenomenon underlying molecular diffusion).  In contrast, in random processes, a Brownian motion (or Wiener process) is a specific stochastic model for continuous-time random walks.  Notably, the random process notion of Brownian motion generates non-differentiable trajectories, which may be viewed as unphysical  (though  still useful nonetheless) \cite{genthon2020,gillespie1996}.  Interestingly, motion models with equivalent properties can be obtained by assuming that the velocity of a particle (i.e., the time-derivative of position) is a continuous-time Gaussian white noise random process \cite{gillespie1996}. Of course, ideal continuous-time white noise processes can also be problematic as they have infinite power, though can still be defined in a generalized sense and are widely used \cite{hajek2015}.

Readers should also be cautioned that the third property of  Brownian motions (i.e., that $w(t)-w(s)$ is a zero-mean Gaussian random variable with variance $\sigma^2|t-s|$) implies that the probability of observing a displacement of $w(t)-w(s)$ between times $s$ and $t$ is independent of the starting position $w(s)$  -- this implies the random process is spatially homogeneous.  While spatial homogeneity assumptions may be convenient from a mathematical perspective, it is clearly not an accurate representation of complicated materials (like biological tissues) that display substantial microstructural heterogeneity -- e.g., look at any electron microscope image of a biological tissue to see how microstructurally-complicated these tissues can be.  Despite not capturing all the nuances of physical diffusion processes in heterogeneous media, the Brownian motion model can still be a useful abstraction.  

Assuming isotropic Brownian motion with diffusion coefficient $D$, the spatial trajectory $\mathbf{x}(t)$ of an individual spin can be modeled as
\begin{equation}
 \mathbf{x}(t) = \mathbf{x}(0) + \sqrt{2 D}\mathbf{w}(t),
\end{equation}
where $\mathbf{x}(0)$ is drawn i.i.d. from some initial probability distribution $p_{\mathrm{init}}(\mathbf{x})$ as described previously and $\mathbf{w}(t) = [w_1(t), w_2(t), w_3(t)]^T$, where $w_1(t)$, $w_2(t)$, and $w_3(t)$ are each assumed to be 1D i.i.d. Brownian motions with $\sigma=1$.

With this choice, it is easily observed that the probability density function for observing a spin displacement of $\mathbf{r}=\mathbf{x}(t) - \mathbf{x}(0)$  after time $t$  is given by
\begin{equation}
P\left( \mathbf{x}(t) - \mathbf{x}(0) = \mathbf{r}\right) = \frac{1}{(4 \pi D t )^{3/2}} e^{-\frac{1}{4Dt} \|\mathbf{r}\|_2^2 }.
\end{equation}
In addition, observe that
\begin{equation}
  E\left[ \mathbf{x}(t) - \mathbf{x}(0) \right] =\mathbf{0},
\end{equation}
\begin{equation}
 E\left[ (\mathbf{x}(t) - \mathbf{x}(0)) (\mathbf{x}(t) - \mathbf{x}(0))^T\right] = \begin{bmatrix} 2 D t & 0 & 0 \\ 0 & 2 D t & 0 \\ 0 & 0 & 2 D t \end{bmatrix},
\end{equation}
and
\begin{equation}
E\left[\|\mathbf{x}(t)-\mathbf{x}(0)\|_2^2 \right] = 6Dt,
\end{equation}
which matches the classical Einstein relation \cite{einstein1905,einstein1956}.  In this case, the displacement probabilities along each axis are i.i.d. Gaussian, with the variance of the distribution determined by the time $t$ and the diffusion coefficient $D$.  Note also that $w_1(t)$, $w_2(t)$, and $w_3(t)$ are Gaussian random processes, and as a result, are completely determined by their mean   and autocorrelation functions \cite{hajek2015} (with $E[\mathbf{w}(t)] = \mathbf{0}$ and $E[\mathbf{w}(s)\mathbf{w}(t)^T] = \min(s,t) \mathbf{I}_3$).     

For a fixed value of $t$, write the accumulated phase as
\begin{equation}
\begin{split}
\phi(t) &= -\gamma  \int_{0}^t \tilde{\mathbf{g}}(\tau) \cdot \mathbf{x}(\tau) d\tau \\
&= -\gamma \int_{0}^t \tilde{\mathbf{g}}(\tau) \cdot \mathbf{x}(0) d\tau + -\gamma  \sqrt{2D} \int_{0}^t \tilde{\mathbf{g}}(\tau) \cdot \mathbf{w}(\tau) d\tau. 
\end{split}
\end{equation}
If we assume that $\int_0^t\tilde{\mathbf{g}}(\tau) d\tau = \mathbf{0}$ at the time $t$ when data is measured (which is generally the case in practical diffusion MR experiments so that a proper gradient echo is formed), the dependence on the initial position $\mathbf{x}(0)$ disappears, and the accumulated phase simplifies to 
\begin{equation}
\begin{split}
\phi(t) &= -\gamma\sqrt{2D} \int_{0}^t \tilde{\mathbf{g}}(\tau) \cdot \mathbf{w}(\tau) d\tau. 
\end{split}
\end{equation}
Because $\mathbf{w}(t)$ is a zero-mean Gaussian random process \cite{hajek2015}, $\phi(t)$ is a Gaussian random variable with  mean
\begin{equation}
 E[\phi(t)] = 0
\end{equation}
and variance
\begin{equation}
\begin{split}
E\left[|\phi(t)|^2\right] &= E\left[ \left(-\gamma \sqrt{2D} \int_0^t \tilde{\mathbf{g}}^T(\tau) \mathbf{w}(\tau) d\tau \right) \left(-\gamma\sqrt{2D} \int_0^t \tilde{\mathbf{g}}^T(s) \mathbf{w}(s) ds \right)^T  \right]\\
 &= 2D \gamma^2 \int_0^t \int_0^t \tilde{\mathbf{g}}^T(\tau) E[\mathbf{w}(\tau) \mathbf{w}^T(s)] \tilde{\mathbf{g}}(s) d\tau ds \\
 &=2 \left(\gamma^2  \int_0^t \int_0^t \tilde{\mathbf{g}}^T(\tau)  \tilde{\mathbf{g}}(s) \min(\tau,s) d\tau ds\right) D\\
 &\triangleq \lambda^2.
 \end{split}
\end{equation}
As a result, we have that Eq.~\eqref{eq:noyesecho} simplifies to 
\begin{equation}
\begin{split}
 N_s \cdot E\left[ e^{-i \gamma \int_{0}^t \tilde{\mathbf{g}}(\tau) \cdot \mathbf{x}(\tau) d\tau} \right] &=  \frac{N_s}{\lambda \sqrt{2\pi }} \int_{-\infty}^\infty e^{i \phi}  e^{-\frac{\phi^2}{2 \lambda^2}} d\phi\\
 &=  N_s e^{- \frac{\lambda^2}{2}}\\
 &= N_s e^{- b D },
 \end{split}
\end{equation}
with 
\begin{equation}
 b \triangleq \gamma^2  \int_0^t \int_0^t \tilde{\mathbf{g}}^T(\tau)  \tilde{\mathbf{g}}(s) \min(\tau,s) d\tau ds.\label{eq:bval}
\end{equation}
We observe that the measured signal decays monoexponentially with respect to the $b$-value and the diffusion coefficient, which matches the classical results \cite{stejskal1965}.  The main potential concern is whether the definition of $b$-value given in Eq.~\eqref{eq:bval} matches the classical definition, as our expression is written in a very different form from the classical one.  We need to simplify!

To make progress, let $\mathbf{k}(t)\triangleq \int_0^t \tilde{\mathbf{g}}(\tau)d\tau$,\footnote{It should be noted that our definition of $\mathbf{k}(t)$ is quite similar to the definition of k-space in MR imaging \cite{liang2000}.  We could have made the definitions even more similar if we had instead defined $\mathbf{k}(t) \triangleq \int_0^t \frac{\gamma}{2\pi} \tilde{\mathbf{g}}(\tau) d\tau$. We have opted not to do that for this primer, but we could have used the alternate definition of $\mathbf{k}(t)$ if we wanted, it would not change the end results.  }  such that $\tilde{\mathbf{g}}(t) = \frac{d}{dt} \mathbf{k}(t)$.  We have $\mathbf{k}(0) = \mathbf{k}(t)=\mathbf{0}$ because we've assumed that $\int_0^t \tilde{\mathbf{g}} (\tau) d \tau = \mathbf{0}$. Similarly, let $\mathbf{h}(t)\triangleq\int_0^t \mathbf{k}(\tau) d\tau$ such that  ${\mathbf{k}}(t) = \frac{d}{dt} \mathbf{h}(t)$.  Then
\begin{equation}
\begin{split}
 b &= \gamma^2  \int_0^t \int_0^t \tilde{\mathbf{g}}^T(\tau)  \tilde{\mathbf{g}}(s) \min(\tau,s) d\tau ds \\
 &= \gamma^2  \int_0^t \left[\int_0^s \tilde{\mathbf{g}}^T(\tau)   \tau d\tau + s\int_s^t \tilde{\mathbf{g}}^T(\tau)   d\tau\right] \tilde{\mathbf{g}}(s) ds \\
 &= \gamma^2  \int_0^t \left[\int_0^s \tilde{\mathbf{g}}^T(\tau)   \tau d\tau + s \left(\mathbf{k}^T(t) - \mathbf{k}^T(s) \right)\right] \tilde{\mathbf{g}}(s) ds\\
 &= \gamma^2  \int_0^t \left[\int_0^s \tilde{\mathbf{g}}^T(\tau)   \tau d\tau  - s  \mathbf{k}^T(s) \right] \tilde{\mathbf{g}}(s) ds.
 \end{split}\label{eq:part}
\end{equation}

Observe that, using integration by parts,  for functions $a(x)$, $b(x) = \frac{d}{dx} a(x)$, and $c(x) = \frac{d}{dx} b(x)$, we have the relation
\begin{equation}
 \int_{x_1}^{x_2}  x c(x) dx = x_2 b(x_2)- x_1 b(x_1) + a(x_1) - a(x_2).
\end{equation}
This allows Eq.~\eqref{eq:part} to be simplified as \begin{equation}
\begin{split}
 b &= \gamma^2  \int_0^t \left[s \mathbf{k}^T(s) + \mathbf{h}^T(0) - \mathbf{h}^T(s)- s \mathbf{k}^T(s) \right] \tilde{\mathbf{g}}(s) ds \\
  &= \gamma^2  \int_0^t \left[\mathbf{h}^T(0) - \mathbf{h}^T(s)\right] \tilde{\mathbf{g}}(s) ds\\
  &= \gamma^2  \mathbf{h}^T(0)\int_0^t\tilde{\mathbf{g}}(s) ds - \gamma^2  \int_0^t\mathbf{h}^T(s)\tilde{\mathbf{g}}(s) ds \\
  &=   - \gamma^2  \int_0^t\mathbf{h}^T(s)\tilde{\mathbf{g}}(s) ds. 
 \end{split}\label{eq:parts2}
\end{equation}
We're very close, and just need to evaluate this last integral.

\begin{figure}
 \centering 
 \includegraphics[width=3in]{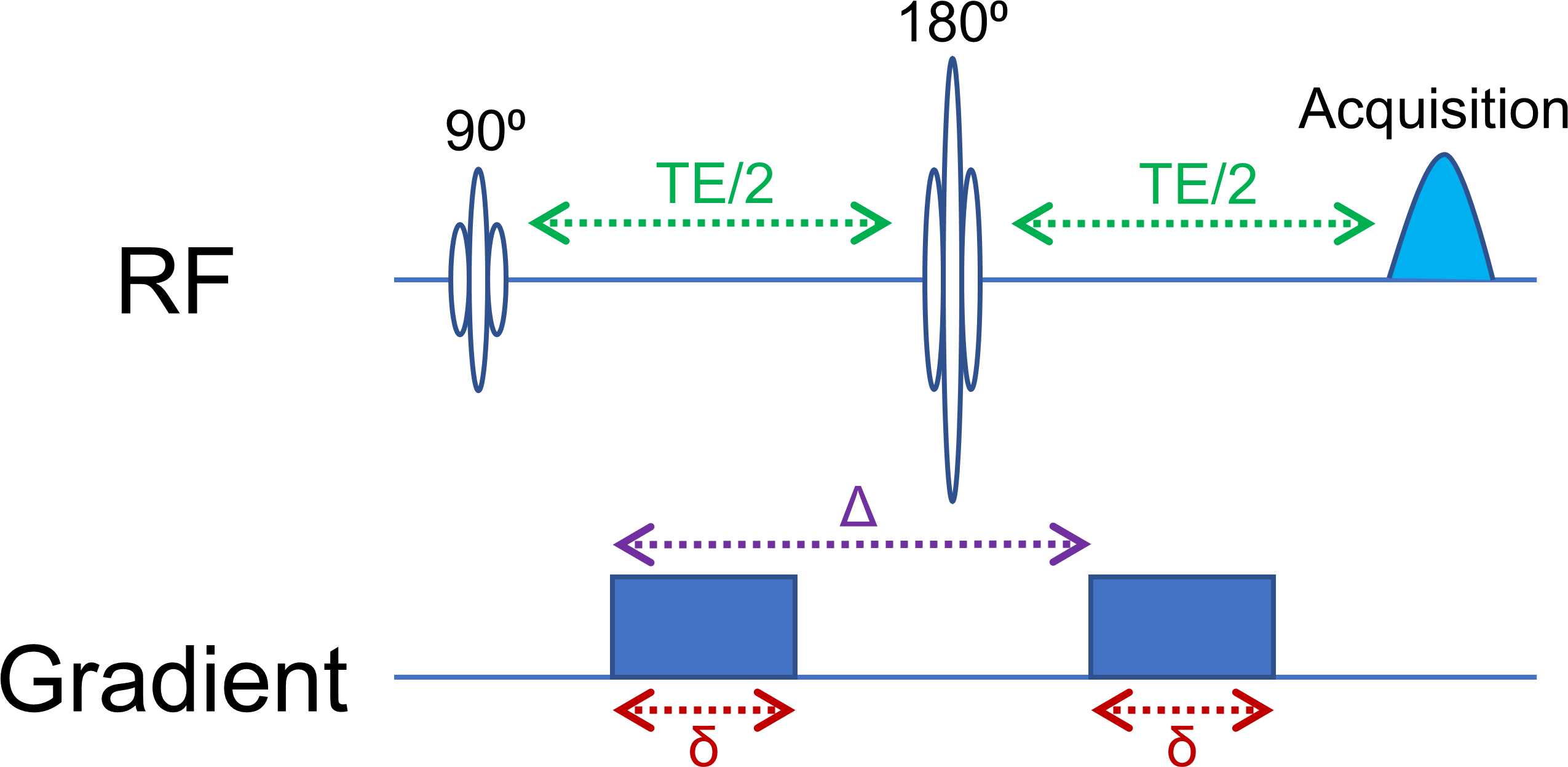}
 \caption{Depiction of the standard Stejskal-Tanner pulsed gradient spin-echo diffusion encoding scheme \cite{stejskal1965}.}
 \label{fig:stej}
\end{figure}

Consider
\begin{equation}
\int_{x_1}^{x_2 } a(x) c(x) dx,
\end{equation}
with $a(x)$, $b(x)$, and $c(x)$ as described previously.  Applying integration by parts, we can write this as 
\begin{equation}
\begin{split}
\int_{x_1}^{x_2 } a(x) c(x) dx &= a(x_2)b(x_2) - a(x_1)b(x_1)  - \int_{x_1}^{x_2 } b^2(x) dx.
\end{split}
\end{equation}
Applying this relation to Eq.~\eqref{eq:parts2} yields
\begin{equation}
\begin{split}
 b &=   - \gamma^2  \int_0^t\mathbf{h}^T(s)\tilde{\mathbf{g}}(s) ds\\
 &=  - \gamma^2 \mathbf{h}^T(t)\mathbf{k}(t) + \gamma^2 \mathbf{h}^T(0)\mathbf{k}(0) + \gamma^2 \int_0^t\mathbf{k}^T(s) \mathbf{k}(s) ds\\
 &=  \gamma^2 \int_0^t\mathbf{k}^T(s) \mathbf{k}(s) ds.
 \end{split}
\end{equation}
This expression matches the conventional results \cite{mattiello1997}!   For example, it is straightforward to derive that the standard Stejskal-Tanner pulsed-gradient spin-echo diffusion encoding scheme (depicted in Fig.~\ref{fig:stej}) with two gradient pulses of duration $\delta$ and magnitude $\sqrt{G_x^2 + G_y^2 + G_z^2}$ that are spaced in time by $\Delta$, the b-value is $b = \gamma^2 (G_x^2 + G_y^2 + G_z^2) \delta^2 (\Delta - \delta/3)$, which matches the standard result \cite{stejskal1965}.  

\subsection{Anisotropic Brownian Motion}
The previous subsection involved isotropic Brownian motion (with no preferred direction), but in biological tissue, diffusion is often observed to be directionally dependent.  An illustration of anisotropic diffusion is given in Fig.~\ref{fig:brown}(b).   A simple form of anisotropy can easily be modeled by appropriate anisotropic scaling and rotation of the i.i.d. Brownian motion $\mathbf{w}(t)$ from the previous subsection:
\begin{equation}
 \mathbf{x}(t) = \mathbf{x}(0) + \sqrt{2} \mathbf{Q} \begin{bmatrix} \sqrt{D_1} & 0 & 0 \\ 0 & \sqrt{D_2} & 0 \\ 0 & 0 & \sqrt{D_3} \end{bmatrix}\mathbf{w}(t),\label{eq:3ani}
\end{equation}
where $\mathbf{Q}$ is a unitary matrix ($\mathbf{Q}^H = \mathbf{Q}^{-1}$) with columns $\mathbf{q}_1$, $\mathbf{q}_2$, and $\mathbf{q}_3$ that define the principal axes of the diffusion process, and $D_1$, $D_2$, and $D_3$ are diffusion coefficients for each orientation (i.e., the diffusion coefficient $D_1$ will be measured along the orientation $\mathbf{q}_1$, etc.).  It is convenient to define the \emph{diffusion tensor} as 
\begin{equation}
 \mathbf{D} \triangleq  \mathbf{Q} \begin{bmatrix} D_1 & 0 & 0 \\ 0 & D_2 & 0 \\ 0 & 0 & D_3 \end{bmatrix} \mathbf{Q}^T.\label{eq:dti}
\end{equation}
The diffusion tensor is symmetric and positive semidefinite.\footnote{Note that, in principle, we could have defined this case using a matrix $\mathbf{Q}$ that is not unitary, with non-orthogonal columns.  In that case, the diffusion tensor can still be defined as in Eq.~\eqref{eq:dti}. However, because the diffusion tensor is symmetric and positive semidefinite, it is well-known from linear algebra that $\mathbf{D}$ will have orthogonal eigenvectors and can always be represented using the spectral decomposition  $\mathbf{D} = \mathbf{P} \boldsymbol{\Lambda} \mathbf{P}^T$, where $\mathbf{P}$ is the unitary matrix of eigenvectors and $\boldsymbol{\Lambda}$ is a diagonal matrix of eigenvectors.  As such, our assumption that $\mathbf{Q}$ is unitary is not restrictive, because for non-unitary matrices, we can always use the eigendecomposition to obtain an equivalent representation involving a unitary matrix.  Similarly, instead of Eq.~\eqref{eq:3ani}, we could have also expressed $\mathbf{x}(t)$ as the weighted sum of more than three i.i.d. Brownian motion processes, e.g., $\mathbf{x}(t) = \mathbf{x}(0) + \sqrt{2} \sum_i \mathbf{q}_i \sqrt{D}_i w_i(t).$   However, due to the characteristics of Gaussian random variables (i.e., sums of Gaussians are still Gaussian), this situation would allow us to  write the diffusion tensor as $\mathbf{D} = \sum_i D_i \mathbf{q}_i \mathbf{q}_i^T$, and then obtain an equivalent representation in terms of three i.i.d. Brownian motions by taking the eigendecomposition of $\mathbf{D}$ and assigning one Brownian motion to each of the three eigenvalue/eigenvector pairs. }

It is easily observed that the probability density function for observing a spin displacement of $\mathbf{r}=\mathbf{x}(t) - \mathbf{x}(0)$  after time $t$  is given in this case by
\begin{equation}
P(\mathbf{x}(t) - \mathbf{x}(0) = \mathbf{r}) = \frac{1}{(4 \pi t )^{3/2} \sqrt{\mathrm{det}(\mathbf{D})}} e^{-\frac{1}{4t} \mathbf{r}^T \mathbf{D}^{-1} \mathbf{r} }.
\end{equation}
We also have that
\begin{equation}
  E\left[ \mathbf{x}(t) - \mathbf{x}(0) \right] =\mathbf{0},
\end{equation}
\begin{equation}
 E\left[ (\mathbf{x}(t) - \mathbf{x}(0)) (\mathbf{x}(t) - \mathbf{x}(0))^T \right] = 2\mathbf{D}t,
\end{equation}
and
\begin{equation}
E\left[\|\mathbf{x}(t)-\mathbf{x}(0)\|_2^2 \right] = 2t \cdot \mathrm{trace}(\mathbf{D}).
\end{equation}
The spin displacement distribution is (again) Gaussian, with the covariance matrix determined by the time $t$ and the diffusion  tensor $\mathbf{D}$.

If we again assume that $\int_0^t\tilde{\mathbf{g}}(\tau) d\tau = \mathbf{0}$, then the accumulated phase simplifies to 
\begin{equation}
\begin{split}
\phi(t) &= -\gamma\sqrt{2} \int_{0}^t \tilde{\mathbf{g}}(\tau)^T \mathbf{Q} \begin{bmatrix} \sqrt{D_1} & 0 & 0 \\ 0 & \sqrt{D_2} & 0 \\ 0 & 0 & \sqrt{D_3} \end{bmatrix} \mathbf{w}(\tau) d\tau. 
\end{split}
\end{equation}
Similar to before, $\phi(t)$ is a zero-mean Gaussian random variable with variance 
\begin{equation}
\begin{split}
E\left[|\phi(t)|^2\right] &= 2 \gamma^2  \int_0^t \int_0^t \tilde{\mathbf{g}}^T(\tau) \mathbf{D}  \tilde{\mathbf{g}}(s) \min(\tau,s) d\tau ds\\
&= 2 \cdot \mathrm{trace}\left(\left( \gamma^2  \int_0^t \int_0^t \tilde{\mathbf{g}}(s)\tilde{\mathbf{g}}^T(\tau) \min(\tau,s) d\tau ds \right) \mathbf{D} \right)\\
 &\triangleq \lambda^2.
 \end{split}
\end{equation}

As a result (and following the same line of arguments from before), we have that Eq.~\eqref{eq:noyesecho} simplifies to:
\begin{equation}
\begin{split}
 N_s \cdot E\left[ e^{-i \gamma \int_{0}^t \tilde{\mathbf{g}}(\tau) \cdot \mathbf{x}(\tau) d\tau} \right]  &=  N_s e^{- \frac{\lambda^2}{2}}\\
 &= N_s e^{- \mathrm{trace}(\mathbf{B} \mathbf{D}) },
 \end{split}
\end{equation}
with 
\begin{equation}
\begin{split}
 \mathbf{B} &\triangleq  \gamma^2  \int_0^t \int_0^t \tilde{\mathbf{g}}(s)\tilde{\mathbf{g}}^T(\tau) \min(\tau,s) d\tau ds\\
 &=  \gamma^2  \int_0^t {\mathbf{k}}(s){\mathbf{k}}^T(s)  ds.
 \end{split}\label{eq:bmat}
\end{equation}
As before, this yields the classical result where the measured signal decays exponentially based on the interaction between the b-matrix $\mathbf{B}$ and the diffusion tensor $\mathbf{D}$ \cite{basser1994a,mattiello1997}. 

Notably, if we take $\mathbf{D} = D \mathbf{I}_3$, then the anisotropic model described in this subsection reduces to the isotropic model described in the previous subsection.  Unsurprisingly, the isotropic Brownian motion model is a special case of the more general anisotropic Brownian motion model.

\subsection{The Diffusion Propagator}
Brownian motion (Wiener process) models can be useful, but they do not capture the true behavior of porous media with boundaries, permeable membranes, or other features that restrict, hinder, or otherwise modify the diffusion process.   For example, if a spin is located very close to an impermeable barrier, then the spin should generally have a higher probability of moving away from the barrier than moving closer to the barrier and should have zero probability of passing through the barrier.  An example of this with impermeable cylindrical boundaries was depicted in Fig.~\ref{fig:brown}(c).  Brownian motion models do not account for such diffusion behavior, as the displacement probability for a Brownian particle  is required to be independent of its starting location. Brownian motion models also do not explain the empirical observation that the apparent diffusion coefficients/tensors measured in a diffusion MR experiment appear to vary substantially as a function of the diffusion time in biological tissues, which is indicative of restrictions to the diffusion process.   The Brownian motion models discussed above also do not model exchange processes, in which a water molecule may move from one compartment to another during its trajectory.

The propagator representation gives one approach to potentially mitigate some of these limitations  \cite{callaghan1991}.  The diffusion propagator $p_\tau(\mathbf{x} | \mathbf{x}_0)$ gives the conditional probability that a spin will be located at position $\mathbf{x}$ after a time delay of $\tau$, conditioned on the  spin being originally located at position $\mathbf{x}_0$.  This is a rich representation, because it allows us to model different types of trajectories for spins located at different spatial locations, which is needed for modeling complicated microstructure configurations.   Another useful representation is the ensemble average propagator (EAP), which is obtained by calculating the probability of observing a  displacement of $\mathbf{r} = \mathbf{x}(\tau) - \mathbf{x}(0)$ after a time delay of $\tau$ by averaging the  propagator over the distribution of initial positions:
\begin{equation}
 p_{\mathrm{EAP}}(\mathbf{r},\tau) = \int p_\tau(\mathbf{x}_0 + \mathbf{r} | \mathbf{x}_0) p_{\mathrm{init}}(\mathbf{x}_0)d\mathbf{x}_0.
\end{equation}

While the propagator and EAP are useful, they notably do not provide a full description of the random process, as they only describe the characteristics at endpoints of the trajectory, and do not provide any information about the actual path that a spin may traverse to get from $\mathbf{x}_0$ to $\mathbf{x}$.  In addition, the propagator representation by itself does not provide any information about the autocorrelation $E[\mathbf{x}(\tau)\mathbf{x}^T(s)]$, which would generally be needed to calculate the signal model corresponding to arbitrary gradient waveforms using Eq.~\eqref{eq:noyesecho}.

However, the propagator representation does lend itself to easy signal modeling when the effective gradient waveform $\tilde{\mathbf{g}}(t)$ has a special sparse form.  This is sometimes called the ``narrow pulse approximation'' \cite{callaghan1991}.  

To give a concrete example, consider a sparse effective gradient waveform comprised of two infinitely-small pulses with opposite sign and separated by a time delay of $\Delta$ such that
\begin{equation}
 \tilde{\mathbf{g}}(t) = \mathbf{p}\left[\delta(t-\Delta) - \delta(t) \right],
\end{equation}
where $\mathbf{p} \in \mathbb{R}^3$ captures the magnitude and orientation of the gradient pulse and $\delta(t)$ represents the Dirac delta function (not to be confused with our previous use of the symbol $\delta$ as the duration of the gradient pulse).  With this very-sparse gradient waveform, we have the simplification that the measured data should be proportional to
\begin{equation}
\begin{split}
 N_s \cdot E\left[ e^{-i \gamma \int_{0}^t \tilde{\mathbf{g}}(\tau) \cdot \mathbf{x}(\tau) d\tau} \right] &=  N_s \cdot E\left[ e^{-i \gamma \mathbf{p} \cdot (\mathbf{x}(\Delta) - \mathbf{x}(0)) } \right]\\
 &=  N_s \int e^{-i \gamma \mathbf{p} \cdot \mathbf{r}} p_{\mathrm{EAP}}(\mathbf{r},\Delta) d\mathbf{r}\\
 &=  N_s \int e^{-i 2\pi \mathbf{q} \cdot \mathbf{r}} p_{\mathrm{EAP}}(\mathbf{r},\Delta) d\mathbf{r},
 \end{split}
\end{equation}
with $q$-space location $\mathbf{q} \triangleq \frac{\gamma}{2\pi} \mathbf{p}$.\footnote{In practice, it is not possible to generate Dirac delta functions, and such gradient pulses are often achieved practically using short rectangular (or trapezoidal) pulses.  In the rectangular case, we would have $ \tilde{\mathbf{g}}(t) = \mathbf{p}\left[\mathrm{rect}\left(\frac{t-\Delta}{\varepsilon}\right) - \mathrm{rect}\left(\frac{t}{\varepsilon}\right) \right]$, where $\varepsilon$ is the width of each pulse.  In this more practical case, we would define the $q$-space position as $\mathbf{q} \triangleq \frac{\gamma}{2\pi} \mathbf{p} \varepsilon$.  In the literature, the parameter $\varepsilon$ is usually denoted using the symbol $\delta$ to be consistent with the Stejskal-Tanner nomenclature given in Fig.~\ref{fig:stej}.  However, we have not adopted that convention here to avoid confusion with the Dirac delta function.}    As a result, we observe that the measured signal is the Fourier transform of the EAP  sampled at position $\mathbf{q}$, which is the standard ``$q$-space model'' of diffusion acquisition \cite{callaghan1991}.  This Fourier representation of data acquisition is quite useful from a signal processing perspective \cite{varadarajan2017}, and e.g., enables nonparametric Fourier reconstruction of the EAP if $q$-space is sampled at the Nyquist rate \cite{wedeen2005}.  

As can be seen, the use of sparse gradients with two infinitely-narrow pulses offers substantial simplifications by abstracting away the random process aspects of the diffusion process, enabling the use of a simpler random variable model.

But what happens with more realistic gradient waveforms where the random process aspects will be more important?  One approach, which has been called the ``impulse-propagator trick'' \cite{codd1999}, approximates the effective gradient waveform as a series of impulses separated by time delays, i.e.,
\begin{equation}
 \tilde{\mathbf{g}}(t) \approx \sum_{n} \mathbf{p}_n \delta(t - nT),
\end{equation}
where $T$ is the time interval between gradient pulses.\footnote{It may be interesting to note that this approximation has much in common with the ``hard-pulse approximation'' used in the completely different context of radiofrequency pulse design \cite{pauly1991}.  And of course, expressions like this (and their limitations) will be familiar to anyone who has studied the digital signal processing/sampling theory concepts related to digital-to-analog conversion.}

Under this model, we can restrict attention to the discrete-time random process $\mathbf{x}_n$ obtained by sampling the continuous-time random process $\mathbf{x}(t)$ such that $\mathbf{x}_n = \mathbf{x}(n T)$.  Assuming that $\mathbf{x}_n$ is  Markovian (i.e., neglecting velocity/momentum effects and/or assuming that $\tau$ is large enough \cite{genthon2020} that the current value of $\mathbf{x}_n$ captures all of the information required to model the distribution of the next sample $\mathbf{x}_{n+1}$, with no need to know the past history of the random process),  the measured data should be proportional to 
\begin{equation}
\begin{split}
 N_s \cdot E\left[ e^{-i \gamma \int_{0}^t \tilde{\mathbf{g}}(\tau) \cdot \mathbf{x}(\tau) d\tau} \right] &=  \iiint \cdots \iiint p_{\mathrm{init}}(\mathbf{x}_0) e^{-i\gamma \mathbf{p}_0\cdot \mathbf{x}_0}  \prod_{n} \left( p_T(\mathbf{x}_n | \mathbf{x}_{n-1}) e^{-i\gamma \mathbf{p}_n\cdot \mathbf{x}_n} \right) d\mathbf{x}_0 d\mathbf{x}_1d\mathbf{x}_2 \cdots.
 \end{split}
\end{equation}
While this expression may appear daunting and cumbersome in general, in practice, there exist convenient representations for the propagators associated with common microstructure geometries (e.g., spheres, cylinders, ellipsoids, etc.) that enable efficient computations \cite{codd1999,henriques2021}.  For tissues containing multiple microstructural environments, the ideas described in the next section (about multi-compartmental modeling) can be applied.   And of course, if the microstructure model is too complicated to calculate analytic expressions, the integrals can also be evaluated numerically using tools like Monte Carlo simulation of spin trajectories.

Double or multiple diffusion encoding experiments \cite{mitra1995,henriques2021} are a notable example of techniques based on this framework.  These approaches are valued for their excellent sensitivity to microstructural characteristics such as  pore size and pore orientation distributions within heterogenous media, which occurs because diffusion-driven spin trajectories can exhibit substantial autocorrelation in the presence of such restrictions.

\subsection{The Compartmental Mixture Model}
In the modern literature, it is quite common to model the signal from a large voxel as a mixture of signals originating from sub-voxel compartments, where each compartment represents a pool of spins with distinct diffusion characteristics.  For example, it can be common to model the signal in brain tissue as originating from a mixture of non-exchanging intracellular, extracellular, and isotropic compartments, each of which has distinct characteristics.  An illustration of a voxel with two distinct compartments was shown in Fig.~\ref{fig:brown}(c).  How could we model this behavior within our random process framework?

To give a concrete illustration, let's say that our voxel contains $M$ distinct non-exchanging subcompartments.  If we choose a spin from the voxel uniformly at random, let's denote the event that we pick a spin from the $m$th subcompartment as $C_m$ which occurs with probability $P(C_m)$ (with $\sum_{m=1}^N P(C_m)=1$).  It is straightforward to see in this case that the measured signal will be a linear mixture of the signals from each compartment.  In particular, applying standard probabilistic reasoning, Eq.~\eqref{eq:noyesecho} becomes
\begin{equation}
\begin{split}
 N_s \cdot E\left[ e^{-i \gamma \int_{0}^t \tilde{\mathbf{g}}(\tau) \cdot \mathbf{x}(\tau) d\tau} \right] =  N_s \sum_{m=1}^M P(C_m) E\left[ e^{-i \gamma \int_{0}^t \tilde{\mathbf{g}}(\tau) \cdot \mathbf{x}(\tau) d\tau} | C_m \right].
 \end{split}
\end{equation}
For the sake of practical tractability, it is often assumed that the characteristics of each compartment are captured by a small number of parameters.  For example, a compartment might be modeled as a spherical pore or a cylindrical pore (e.g., parameterized by the diffusion coefficient corresponding to free/unrestricted diffusion, the pore radius, and surface relaxation parameters).  An even simpler approach might be to just assume that all compartments follow Brownian motion (Wiener process) models, leading to a simple multi-tensor model
\begin{equation}
\begin{split}
 N_s \cdot E\left[ e^{-i \gamma \int_{0}^t \tilde{\mathbf{g}}(\tau) \cdot \mathbf{x}(\tau) d\tau} \right] =  N_s \sum_{m=1}^M P(C_m) e^{-\mathrm{trace}(\mathbf{B} \mathbf{D}_m)}.
 \end{split}
\end{equation} 

Generalizing these ideas, it is also possible to describe diffusion using ``spectral''/continuum models\footnote{The term ``spectral'' is used in this context to refer to a continuous distribution, with analogy to the way that a voxel is modeled as a continuous distribution of different frequency components in conventional MR spectroscopy.  However, readers are cautioned  not to confuse this kind of spectral modeling with concepts related to the power spectral density of a random process, which we will discuss shortly in the sequel.} that assume an infinite number of compartments within a single voxel.  For example, assuming an infinite mixture of diffusion tensors and letting $P(\mathbf{D})$ denote the corresponding distribution function (or spectrum), one could model the signal from a voxel as \cite{yablonskiy2010,wang2011b}\footnote{Interestingly, some of the recent literature has ascribed special significance to the ``q-space trajectory''  used to achieve diffusion weighting when working with a (spectral) diffusion tensor distribution model.  This may be confusing to some readers of this primer, as our signal expression in Eq.~\eqref{eq:dtd} seems to only depend on the matrix $\mathbf{B}$ and the diffusion tensor distribution $P(\mathbf{D})$, with no apparent dependence on any kind of trajectory.  To clarify, the ``q-space trajectory'' considered in such work is simply $\gamma\mathbf{k}(t)$ in our notation, and is related to the b-matrix $\mathbf{B}$ as previously described in Eq.~\eqref{eq:bmat}.  Put another way, the matrix $\mathbf{B}$ (which, due to symmetry, can be specified using only 6 real numbers) fully describes  the only characteristics of the q-space trajectory that are relevant for modeling the signal from a diffusion tensor distribution model.  Notably, there are many inconsequential features of the q-space trajectory that are not captured by $\mathbf{B}$, and there will generally be infinitely many q-space trajectories that yield a given $\mathbf{B}$ matrix.  As such, the detailed temporal characteristics of the q-space trajectory might be viewed as having lesser importance in the context of diffusion tensor distribution models.  However, as we will see in the sequel, the temporal characteristics (or more precisely, the temporal-frequency characteristics) of the q-space trajectory   can become much more important when assuming random process models with more complicated autocorrelation structure.}
\begin{equation}
  N_s \cdot E\left[ e^{-i \gamma \int_{0}^t \tilde{\mathbf{g}}(\tau) \cdot \mathbf{x}(\tau) d\tau} \right] = N_s \int P(\mathbf{D}) e^{-\mathrm{trace}(\mathbf{B} \mathbf{D})} d\mathbf{D}.\label{eq:dtd}
\end{equation}
Clearly, the previous $M$-compartment model can be obtained as a special case by assuming that $P(\mathbf{D})$ is comprised of $M$ Dirac delta functions.

Note that while diffusion tensor models are commonly used with spectral representations,  there is also nothing special about diffusion tensors that make them more or less suited for spectral representation compared to other signal models.  For instance, Ref.~\cite{assaf2008} describes a method in which a voxel is modeled as a mixture of cylindrical compartments with a spectrum of different radius parameters \cite{assaf2008}.

These kinds of spectral representations are substantially more flexible and are also likely more accurate representations of  physical reality  than  compartmental models with a small  number of discrete compartments. However, the typical exponential-decay characteristics of diffusion contrast combined with the fact that the spectrum is continuous (infinite dimensional!) can lead to an ill-posed problem, which means that it can be difficult to estimate the spectrum from measured data unless additional assumptions are made.  See Refs.~\cite{kim2017a,kim2020,slator2021} for further discussion of such ill-posedness, as well as for discussion of higher-dimensional acquisition strategies that supplement diffusion encoding with other contrast encoding mechanisms to help mitigate such problems.

\subsection{Spectral Density Modeling}
Another model for the diffusion process can be obtained by assuming that the velocity $\mathbf{v}(t) = \frac{d}{dt} \mathbf{x}(t)$ of each spin is a zero-mean wide-sense stationary Gaussian random process with autocorrelation $E[\mathbf{v}(t)\mathbf{v}(s)^T] = \mathbf{R}_\mathbf{v}(t-s)$ \cite{stepisnik1981,gore2010}.  Note that this autocorrelation is matrix-valued (i.e., $\mathbf{R}_\mathbf{v}(\tau): \mathbb{R} \rightarrow \mathbb{R}^{3\times 3}$).

Let
\begin{equation}
2\mathbf{D}(\Omega) \triangleq \int_{-\infty}^\infty e^{-i 2 \pi \tau \Omega} \mathbf{R}_\mathbf{v}(\tau) d\tau
\end{equation}
denote the power spectral density of the velocity, and assume that 
\begin{equation}
 \mathbf{x}(t) = \mathbf{x}(0) + \int_0^t \mathbf{v}(\tau)d\tau,
\end{equation}
where the distribution of the velocity $\mathbf{v}(t)$ is assumed to be independent of the starting position $ \mathbf{x}(0)$.

Under this model, the accumulated phase can be written as
\begin{equation}
\phi(t) = -\gamma \int_{0}^t \tilde{\mathbf{g}}(\tau)^T \left(\mathbf{x}(0) + \int_0^\tau \mathbf{v}(\alpha)d\alpha \right) d\tau.
\end{equation}
If we again assume that $\int_0^t\tilde{\mathbf{g}}(\tau) d\tau = \mathbf{0}$, then the starting point is irrelevant, and this simplifies to 
\begin{equation}
\phi(t) = -\gamma \int_{0}^t   \int_0^\tau \tilde{\mathbf{g}}(\tau)^T\mathbf{v}(\alpha)d\alpha  d\tau.
\end{equation}
Because we had assumed that $\mathbf{v}(t)$ was a zero-mean Gaussian random process, the accumulated phase will also be zero-mean Gaussian, and it remains to calculate its variance.   Observe that
\begin{equation}
\begin{split}
 E\left[|\phi(t)|^2 \right] &= \gamma^2 E\left[\left(\int_{0}^t   \int_0^\tau \tilde{\mathbf{g}}(\tau)^T\mathbf{v}(\alpha)d\alpha  d\tau\right)\left(\int_{0}^t   \int_0^\nu \mathbf{v}(\beta)^T\tilde{\mathbf{g}}(\nu)d\beta  d\nu\right) \right] \\
 &= \gamma^2 \int_{0}^t   \int_0^\tau \int_{0}^t   \int_0^\nu \tilde{\mathbf{g}}(\tau)^TE\left[\mathbf{v}(\alpha)  \mathbf{v}(\beta)^T\right]\tilde{\mathbf{g}}(\nu) d\beta  d\nu d\alpha  d\tau\\
 &= \gamma^2 \int_{0}^t   \int_0^\tau \int_{0}^t   \int_0^\nu \tilde{\mathbf{g}}(\tau)^T \mathbf{R}_\mathbf{v}(\alpha-\beta)\tilde{\mathbf{g}}(\nu)d\beta  d\nu d\alpha  d\tau  \\
 &= \gamma^2 \int_{0}^t   \int_{0}^t    \mathrm{trace}\left(\tilde{\mathbf{g}}(\nu)\tilde{\mathbf{g}}(\tau)^T \left( \int_0^\tau \int_0^\nu \mathbf{R}_\mathbf{v}(\alpha-\beta) d\beta d\alpha\right)\right)  d\tau   d\nu.
 \end{split}
\end{equation}
Defining $\mathbf{k}(t)$ as before and applying integration by parts, this expression becomes
\begin{equation}
\begin{split}
 E\left[|\phi(t)|^2 \right] &=  \gamma^2 \int_{0}^t   \int_{0}^t    \mathrm{trace}\left(\tilde{\mathbf{g}}(\nu)\tilde{\mathbf{g}}(\tau)^T \left( \int_0^\tau \int_0^\nu \mathbf{R}_\mathbf{v}(\alpha-\beta) d\beta d\alpha\right)\right)  d\tau   d\nu\\
 &= -\gamma^2 \int_0^t \int_0^t \mathrm{trace}\left(\tilde{\mathbf{g}}(\nu) \mathbf{k}(\tau)^T  \left(\int_0^\nu \mathbf{R}_\mathbf{v}(\tau-\beta)d\beta \right)\right)  d\tau d\nu \\ 
 &= \gamma^2 \int_0^t \int_0^t \mathrm{trace}\left(\mathbf{k}(\nu) \mathbf{k}(\tau)^T   \mathbf{R}_\mathbf{v}(\tau-\nu) \right)  d\tau d\nu\\
 &= \gamma^2 \mathrm{trace}\left(\int_0^t \int_0^t \mathbf{k}(\nu) \mathbf{k}(\tau)^T   \mathbf{R}_\mathbf{v}(\tau-\nu) d\tau d\nu\right). 
 \end{split} \label{eq:spec}
\end{equation}
An interesting observation is that if $\mathbf{R}_\mathbf{v}(\tau - \nu) = 2\mathbf{D} \delta(\tau-\nu)$ such that the power spectral density satisfies $\mathbf{D}(\Omega) = \mathbf{D}$ for some positive semidefinite diffusion tensor $\mathbf{D}$ (i.e., if the velocity $\mathbf{v}(t)$ is a Gaussian white noise random process \cite{hajek2015}), then we obtain the exact same signal model as previously obtained in the case of anisotropic Brownian motion. This is entirely consistent with our previous comments about a Gaussian white noise model for the velocity leading to the same results as the Brownian motion (Wiener process) model for the position \cite{gillespie1996}!  

But what if the velocity autocorrelation has more complicated structure?  Let 
\begin{equation}
 \mathbf{z}(\Omega) \triangleq \int_0^t e^{-i 2\pi \tau \Omega} \mathbf{k}(\tau) d\tau
\end{equation}
be the frequency spectrum of the $\mathbf{k}(t)$ trajectory.  Then, by applying the convolution and time-reversal properties of the Fourier transform, Eq.~\eqref{eq:spec} reduces to
\begin{equation}
\begin{split}
 E\left[|\phi(t)|^2 \right] &=  2\gamma^2 \mathrm{trace}\left(\int_{-\infty}^\infty \int_0^t e^{i 2\pi \tau \Omega} \mathbf{z}(\Omega) \mathbf{k}(\tau)^T   \mathbf{D}(\Omega) d\tau d\Omega \right)\\
 &= 2\gamma^2 \mathrm{trace}\left(\int_{-\infty}^\infty  \mathbf{z}(\Omega) \mathbf{z}(\Omega)^H   \mathbf{D}(\Omega)  d\Omega \right)\\
 &\triangleq \lambda^2.
 \end{split}
\end{equation}
Following previous arguments, the measured signal will thus be proportional to
\begin{equation}
\begin{split}
 N_s \cdot E\left[ e^{-i \gamma \int_{0}^t \tilde{\mathbf{g}}(\tau) \cdot \mathbf{x}(\tau) d\tau} \right]  &=  N_s e^{- \frac{\lambda^2}{2}}\\
 &= N_s e^{- \mathrm{trace}\left(\int_{-\infty}^\infty \mathbf{B}(\Omega) \mathbf{D}(\Omega)d\Omega\right) },
 \end{split}
\end{equation}
with 
\begin{equation}
 \mathbf{B}(\Omega) \triangleq \gamma^2 \mathbf{z}(\Omega)  \mathbf{z}(\Omega)^H.
\end{equation}

These results suggest that, by appropriate design of the effective diffusion encoding gradient waveform $\tilde{\mathbf{g}}(t)$, we can sensitize our experimental measurements to different components of the power spectral density $\mathbf{D}(\Omega)$.  Oscillating gradient methods \cite{gore2010} represent an important class of such methods, for which the use of sinusoidal gradient waveforms enables the targeted probing of specific frequencies of interest $\Omega$.  The ability to estimate $\mathbf{D}(\Omega)$ is potentially valuable because different microstructural geometries are known to exhibit different spectral characteristics  \cite{gore2010}.  Of course, it should still be kept in mind that this signal model is based on the assumption that the spin velocity is a wide-sense stationary Gaussian random process.  This assumption may be reasonable in certain contexts, though is not entirely general and is definitely not a perfect model for diffusion within microstructurally-complex media.

\section{Final Thoughts}
This primer described theoretical models for the measured diffusion signal in a diffusion-encoded MR experiment from a random process point of view. Although none of the results we obtained are new, we expect that  our derivations (based on concepts from random processes and signal processing with explicitly-stated assumptions) may be more intuitive to some readers.

Throughout this description, we have emphasized repeatedly that many of the results are based on modeling assumptions that are imperfect in various ways.  These comments are not meant to imply that these models are not useful because of their imperfections, and were instead included for the sake of pedagogy and to encourage critical thinking. Indeed, a common aphorism  in statistics (usually attributed to the statistician George E. P. Box) is that ``all models are wrong but some are useful.''   For example, in the context of diffusion MR experiments, even very simple models like the Brownian motion model can yield apparent diffusion coefficients $D$ and apparent diffusion tensors $\mathbf{D}$ that are still sensitive to microstructural changes in biological tissues and can be useful as biomarkers for various disease processes.  A related point is that, if a model happens to produce accurate predictions of experimental data, it does not necessarily mean that the model was correct. In practice, there can be many distinct physics models that can yield identical or nearly identical measurements.  Modeling and interpreting data requires critical thinking  -- if you are just calculating and interpreting numbers without thinking carefully about where those numbers came from, you are likely doing something wrong!

\addcontentsline{toc}{section}{\refname}
\bibliographystyle{IEEEtran}
\bibliography{./reference}

\end{document}